\def\apj{{ApJ}}

\def\mnras{{MNRAS}}

\def\deg{$^\circ$}
\def\kms{$\,{\rm km}\,{\rm s}^{-1}$}
\def\Mpc{$\,h^{-1}\,{\rm Mpc}$}
\def\invMpc{$\,h\,{\rm Mpc}^{-1}$}
\def\etal{et~al.}

\documentclass[cup5b]{caps}
\usepackage{graphicx}
\usepackage{amssymb}
\usepackage{ociwsymp2}
\HeadText{M. Colless}

\begin{document}

\pagenumbering{arabic}

\author[]{MATTHEW COLLESS \\ 
Research School of Astronomy and Astrophysics \\
The Australian National University}

\chapter{Cosmological Results from the \\ 2dF Galaxy Redshift Survey}

\begin{abstract}
The 2dF Galaxy Redshift Survey (2dFGRS) has produced a three-dimensional
map of the distribution of 221,000 galaxies covering 5\% of the sky and
reaching out to a redshift $z\approx 0.3$. This is first map of the
large-scale structure in the local Universe to probe a statistically
representative volume, and provides direct evidence that the large-scale
structure of the Universe grew through gravitational instability.
Measurements of the correlation function and power spectrum of the
galaxy distribution have provided precise measurements of the mean mass
density of the Universe and the relative contributions of cold dark
matter, baryons, and neutrinos. The survey has produced the first
measurements of the galaxy bias parameter and its variation with galaxy
luminosity and type. Joint analysis of the 2dFGRS and cosmic microwave 
background power spectra
gives independent new estimates for the Hubble constant and the vacuum
energy density, and constrains the equation of state of the vacuum.
\end{abstract}

\section{Introduction}
\label{2dFGRS_intro}

The 2dF Galaxy Redshift Survey (2dFGRS) was made possible by the
2-degree Field (2dF) fiber spectrograph, which was specifically
conceived as a tool for performing a massive redshift survey to
precisely measure fundamental cosmological parameters. The
state-of-the-art redshift surveys of the early 1990's, such as the Las
Campanas Redshift Survey (Shectman \etal\ 1996) and the {\it IRAS}\ Point
Source Catalog redshift survey (Saunders \etal\ 2000), either did not
cover sufficiently large volumes to be statistically representative of
the large-scale structure, or covered large volumes too sparsely to
provide precise measurements. An order-of-magnitude increase in the
survey volume and sample size was needed to enter the regime of
``precision cosmology,'' and this became the foundational goal of the
2dFGRS.

The 2dF spectrograph can observe 400 objects simultaneously over a
2\deg-diameter field of view (Taylor \& Gray 1990; Lewis \etal\ 2002a),
and was first placed on the Anglo-Australian Telescope (AAT) in
November 1995. The first spectra were taken in mid-1996, and scheduled
observations with 2dF at full functionality began in September 1997. The
first major redshift survey observing run occurred in October 1997, with
the survey passing 50,000 redshifts in mid-1999 (Colless 1999) and 100,000
redshifts in mid-2000. The first 100,000 redshifts and spectra were
released publicly in June 2001 (Colless et al. 2001), and the
200,000-redshift mark was achieved toward the end of 2001. The survey
observations were completed in April 2002, after 5 years and 272 nights
on the AAT. The final survey is an order of magnitude larger than any
previous redshift survey, and comparable to the ongoing redshift survey
of the Sloan Digital Sky Survey (Bernardi, this volume).

The source catalog for the 2dFGRS was a revised and extended version
of the APM galaxy catalog (Maddox \etal\ 1990), which was created by
scanning the photographic plates of the UK Schmidt Telescope Southern
Sky Survey. The survey targets were chosen to be galaxies with
extinction-corrected magnitudes brighter than $b_J$ = 19.45 mag. The galaxies
were distinguished from stars by the APM image classification algorithm
described by Maddox \etal, conservatively tuned to include all galaxies
at the expense of also including a 5\% contamination by stars.

\begin{figure}
%\centering
\hspace{-1.0cm}
\leavevmode
\includegraphics[clip,height=1.05\columnwidth,angle=270]{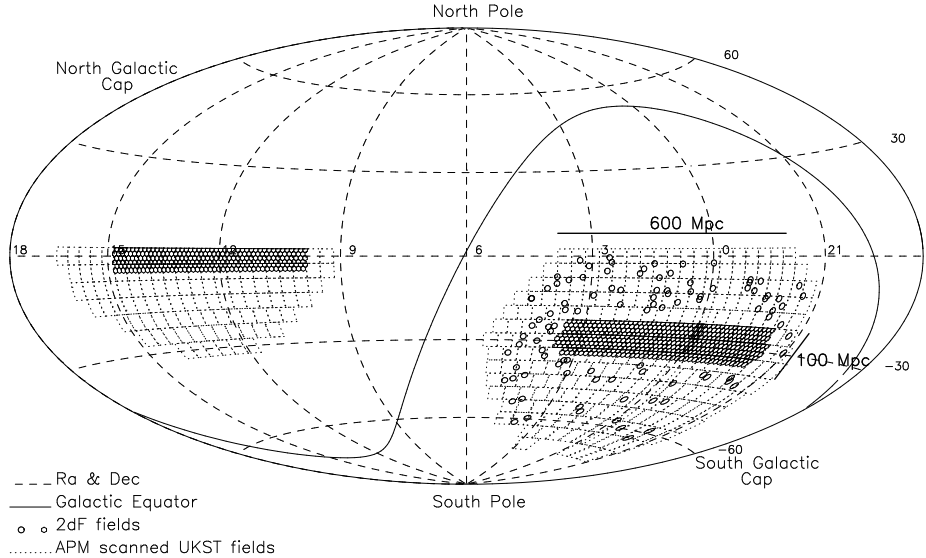}
\caption{A map of the sky showing the locations of the two 2dFGRS survey
strips (NGP strip at left, SGP strip at right) and the random fields.
Each 2dF field in the survey is shown as a small circle; the sky survey
plates from which the source catalog was constructed are shown as
dotted squares. The scale of the strips at the mean redshift of the
survey is indicated.}
\label{2dFGRS_regions}
\end{figure}

The main survey regions were two declination strips, one in the southern
Galactic hemisphere spanning 80\deg$\times$15\deg\ around the South
Galactic Pole (the SGP strip), and the other in the northern Galactic
hemisphere spanning 75\deg$\times$10\deg\ along the celestial equator
(the NGP strip); in addition, there were 99 individual 2dF ``random''
fields spread over the southern Galactic cap (see
Fig.~\ref{2dFGRS_regions}). The large volume that is sparsely probed
by the random fields allows the survey to measure structure on scales
greater than would be permitted by the relatively narrow widths of the
main survey strips. In total, the survey covers approximately 1800 deg$^2$,
and has a median redshift depth of $z$ = 0.11. An adaptive tiling
algorithm was used to optimally place the 900 2dF fields over the survey
regions, giving a highly complete and uniform sample of galaxies on the
sky.

Redshifts were measured from 2dF spectra that covered the range from
3600 \AA\ to 8000 \AA\ at a resolution of 9.0 \AA. Redshift measurements
were obtained both from cross-correlation with template spectra and from
fitting emission lines. All redshifts were visually checked and assigned
a quality parameter $Q$ in the range 1--5; accepted redshifts
($Q$$\ge$3) were found to be 98\% reliable and to have a typical
uncertainty of 85 \kms. The overall redshift completeness for accepted
redshifts was 92\%, although this varied with magnitude. The variation
in the redshift completeness with position and magnitude is fully
accounted for by the survey completeness mask (Colless \etal\ 2001;
Norberg \etal\ 2002b). 

\begin{figure}
%\centering
\hspace{-1.0cm}
\leavevmode
\includegraphics[clip,width=1.05\columnwidth]{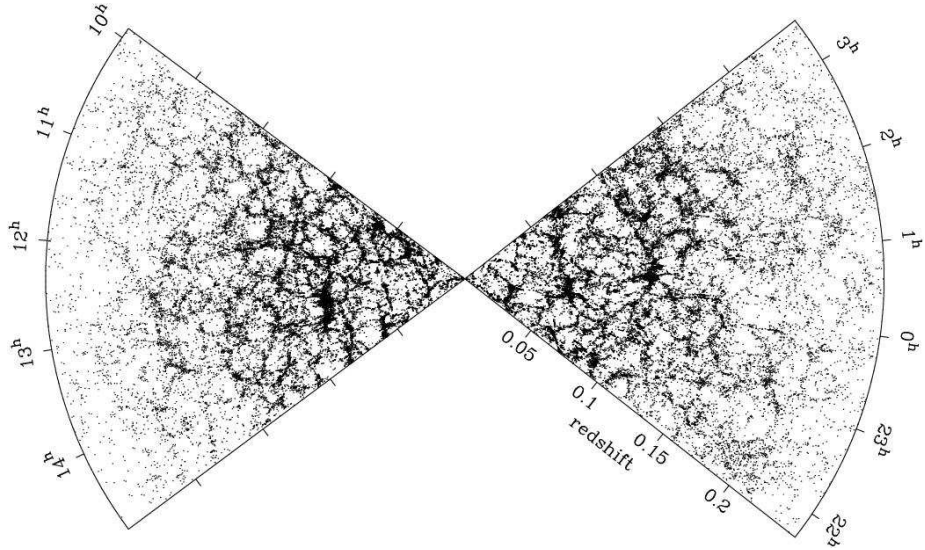}
\caption{The large-scale structures in the galaxy distribution are shown
in this 3\deg-thick slice through the 2dFGRS map.
The slice cuts through the NGP strip (at left) and the SGP strip (at
right), and contains 63,000 galaxies.}
\label{2dFGRS_slice}
\end{figure}

Figure~\ref{2dFGRS_slice} shows a thin slice through the
three-dimensional map of over 221,000 galaxies produced by the 2dFGRS.
This 3\deg-thick slice passes through both the NGP strip (at left) and
the SGP strip (at right). The decrease in the number of galaxies toward
higher redshifts is an effect of the survey selection by
magnitude --- only intrinsically more luminous galaxies are brighter than
the survey magnitude limit at higher redshifts. The clusters, filaments,
sheets and voids making up the large-scale structures in the galaxy
distribution are clearly resolved. The fact that there are many such
structures visible in the figure is a qualitative demonstration that the
survey volume comprises a representative sample of the Universe; the
small amplitude of the density fluctuations on large scales is
quantified by the power spectrum, as discussed in the next section.

\section{The Large-scale Structure of the Galaxy Distribution}
\label{2dFGRS_lss}

In cosmological models where the initial density fluctuations form a
Gaussian random field, such as most inflationary models, the large-scale
structure of the galaxy distribution in the linear regime is completely
characterized in statistical terms by just two quantities: the mean
density and the rms fluctuations in the density as a function of scale.
The latter are quantified either through the two-point correlation
function or the power spectrum, which are Fourier transforms of each
other. However, a redshift survey does not determine the real-space
positions of the galaxies, but rather the redshift-space positions,
where the line-of-sight component is not the distance to the galaxy but
the galaxy's velocity. This velocity is the combination of the Hubble
velocity (which {\em is} directly related to the distance) and the
galaxy's peculiar velocity (the motion produced by the gravitational
attraction of the local mass distribution).

\begin{figure}
%\centering
\hspace{-1.0cm}
\leavevmode
\includegraphics[clip,width=.56\columnwidth]{2dFGRS_stats_pk.eps} \hfil
\includegraphics[clip,width=.50\columnwidth]{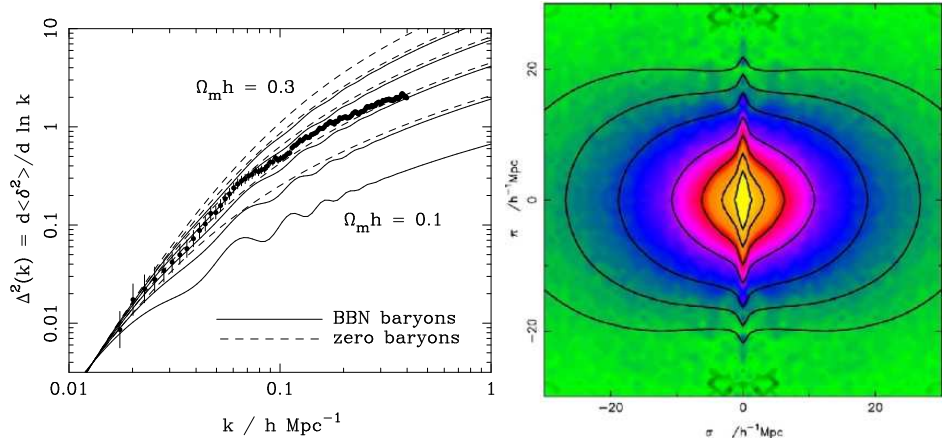}
\caption{Large-scale structure statistics from the 2dFGRS. The left
panel shows the dimensionless power spectrum $\Delta^2(k)$ (Percival
\etal\ 2001; Peacock \etal\ 2003). Overlaid are the predicted
linear-theory CDM power spectra with shape parameters $\Omega h$ = 0.1,
0.15, 0.2, 0.25, and 0.3, with the baryon fraction predicted by Big Bang
nucleosynthesis (solid curves) and with zero baryons (dashed curves).
The right panel shows the two-dimensional galaxy correlation function,
$\xi(\sigma,\pi)$, where $\sigma$ is the separation across the line of
sight and $\pi$ is the separation along the line of sight (Hawkins
\etal\ 2003). The greyscale image is the observed $\xi(\sigma,\pi)$, and
the contours show the best-fitting model.}
\label{2dFGRS_stats}
\end{figure}

The statistical properties of the large-scale structure of the galaxy
distribution observed in redshift space are summarized in
Figure~\ref{2dFGRS_stats}, which shows both the correlation function and
the power spectrum obtained from the 2dFGRS. The structure on very large
scales (several tens to hundreds of Mpc) is best represented by the
power spectrum; on smaller scales, where peculiar velocities become more
significant and the shape of the power spectrum (as well as the
amplitude) differs between redshift space and real space, the
redshift-space structure is most clearly shown in the two-dimensional
correlation function (see \S\ref{2dFGRS_dist} below).

The power spectrum, shown in the left panel of
Figure~\ref{2dFGRS_stats}, is well determined from the 2dFGRS on scales
less than about 400\Mpc\ (wavenumbers $k>0.015$), and its shape is
little affected by nonlinear evolution of the galaxy distribution on
scales greater than about 40\Mpc\ ($k<0.15$). Over this decade in scale,
the power spectrum is well fitted by a cold dark matter (CDM) model
having a shape parameter $\Gamma = \Omega_m h = 0.20 \pm 0.03$ (Percival
\etal\ 2001). For a Hubble constant around $70\, {\rm km}\,{\rm
s}^{-1}\,{\rm Mpc}^{-1}$ (i.e., $h \approx 0.7$), this implies a mean
mass density $\Omega_m \approx 0.3$. The power spectrum also shows some
evidence for acoustic oscillations produced by baryon-photon coupling in
the early Universe (see \S\ref{2dFGRS_omega}).

The right panel of Figure~\ref{2dFGRS_stats} shows the redshift-space
two-point correlations as a function of the separations along and across
the line of sight, and reveals two main deviations from circular
symmetry due to peculiar velocity effects. On intermediate scales, for
transverse separations of a few tens of Mpc, the contours of the
correlation function are flattened along the line of sight due to the
coherent infall of galaxies as structures form in the linear regime. The
detection of this effect in the 2dFGRS is a clear confirmation that
large-scale structure grows by the gravitational amplification of
density fluctuations (Peacock \etal\ 2001), and allows a direct
measurement of the mean mass density of the Universe (see
\S\ref{2dFGRS_omega}). The other effect is the stretching of the contours
along the line of sight at small transverse separations. This is the
finger-of-God effect due to the large peculiar velocities of collapsed
structures in the nonlinear regime.

\section{The Bias of the Galaxy Distribution}
\label{2dFGRS_bias}

A fundamental issue in employing redshift surveys of galaxies as probes
of cosmology is the relationship between the observed galaxy
distribution and the underlying mass distribution, which is what
cosmological models most directly predict. Some bias of the galaxies
with respect to the mass is expected on theoretical grounds, but the
nature and extent of the effect was not previously well determined. The
large size of the 2dFGRS has allowed a thorough investigation of this
question. 

The simplest model for galaxy biasing postulates a linear relation
between fluctuations in the galaxy distribution and fluctuations in the
mass distribution: $\delta n/n = b \delta \rho/\rho$. In this case the
galaxy power spectrum is related to the mass power spectrum by $P_g(k) =
b^2 P_m(k)$. Such a relationship is expected to hold in the linear
regime (up to stochastic variations). The first-order relationship
between galaxies and mass can therefore be determined by comparing the
measured galaxy power spectrum to the matter power spectrum based on a
model fit to the cosmic microwave 
background (CMB) power spectrum, linearly evolved to $z=0$ and
extrapolated to the smaller scales covered by the 2dFGRS power spectrum.
Applying this approach, Lahav \etal\ (2002) find that the linear bias
parameter for an $L^*$ galaxy at zero redshift is $b(L^*,z=0) = (0.96
\pm 0.08) \exp[-\tau+0.5(n-1)]$, where $\tau$ is the optical depth due
to reionization and $n$ is the spectral index of the primordial mass
power spectrum.

An alternative way of determining the bias employs the higher-order
correlations between galaxies in the intermediate, quasi-linear regime.
The higher-order correlations are generated by nonlinear gravitational
collapse, and so depend on the clustering of the dominant dark matter
rather than the galaxies. Thus the stronger the higher-order clustering,
the higher the dark matter normalization, and the lower the bias. An
analysis of the bispectrum (the Fourier transform of the three-point
correlation function) by Verde \etal\ (2002) yields $b(L^*,z=0) = 0.92
\pm 0.11$, a result based solely on the 2dFGRS. Moreover, including a
second-order quadratic bias term does not improve the fit of the bias
model to the observed bispectrum.

\begin{figure}
%\centering
\hspace{-1.0cm}
\leavevmode
\includegraphics[clip,width=.56\columnwidth]{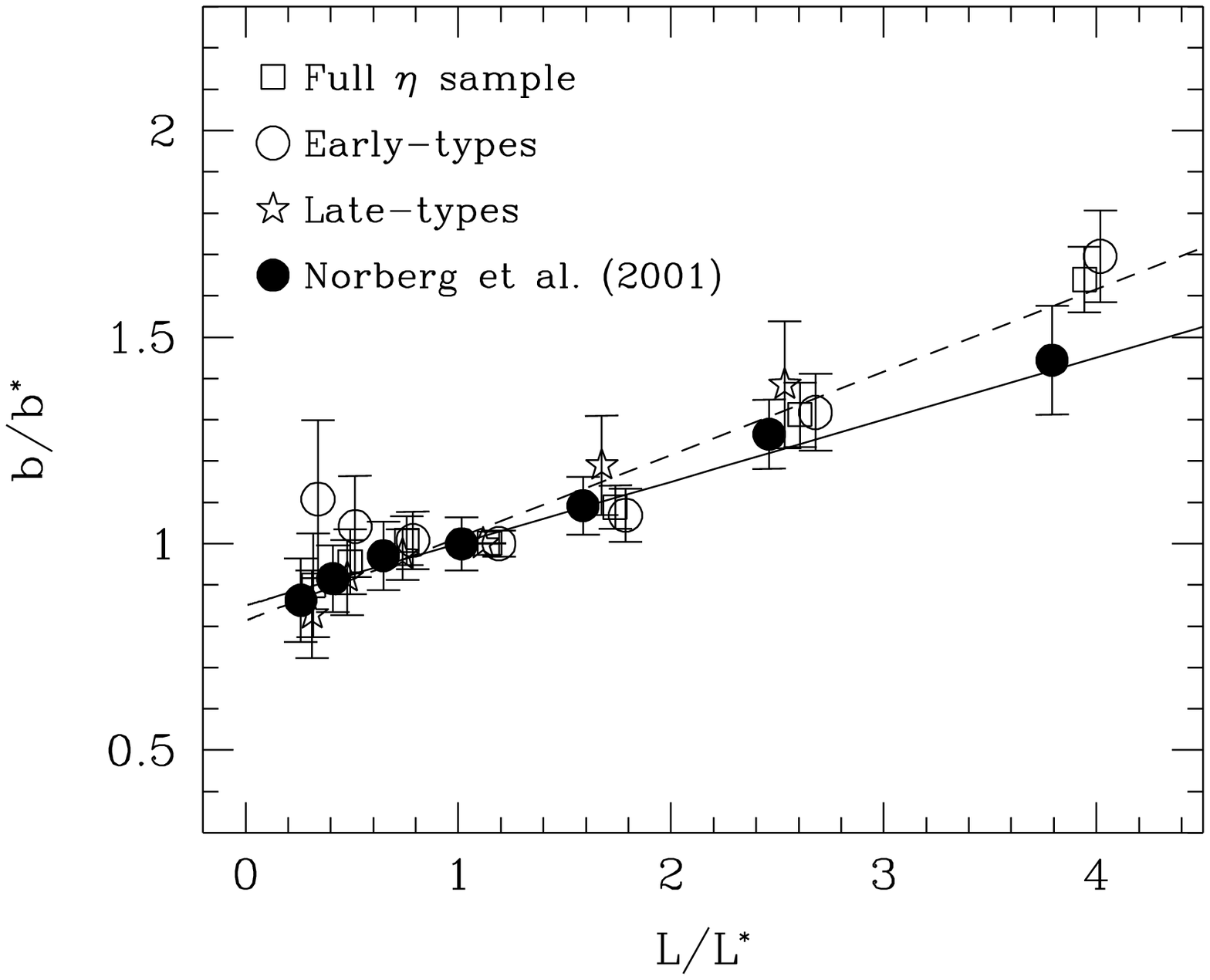} \hfil
\includegraphics[clip,width=.50\columnwidth]{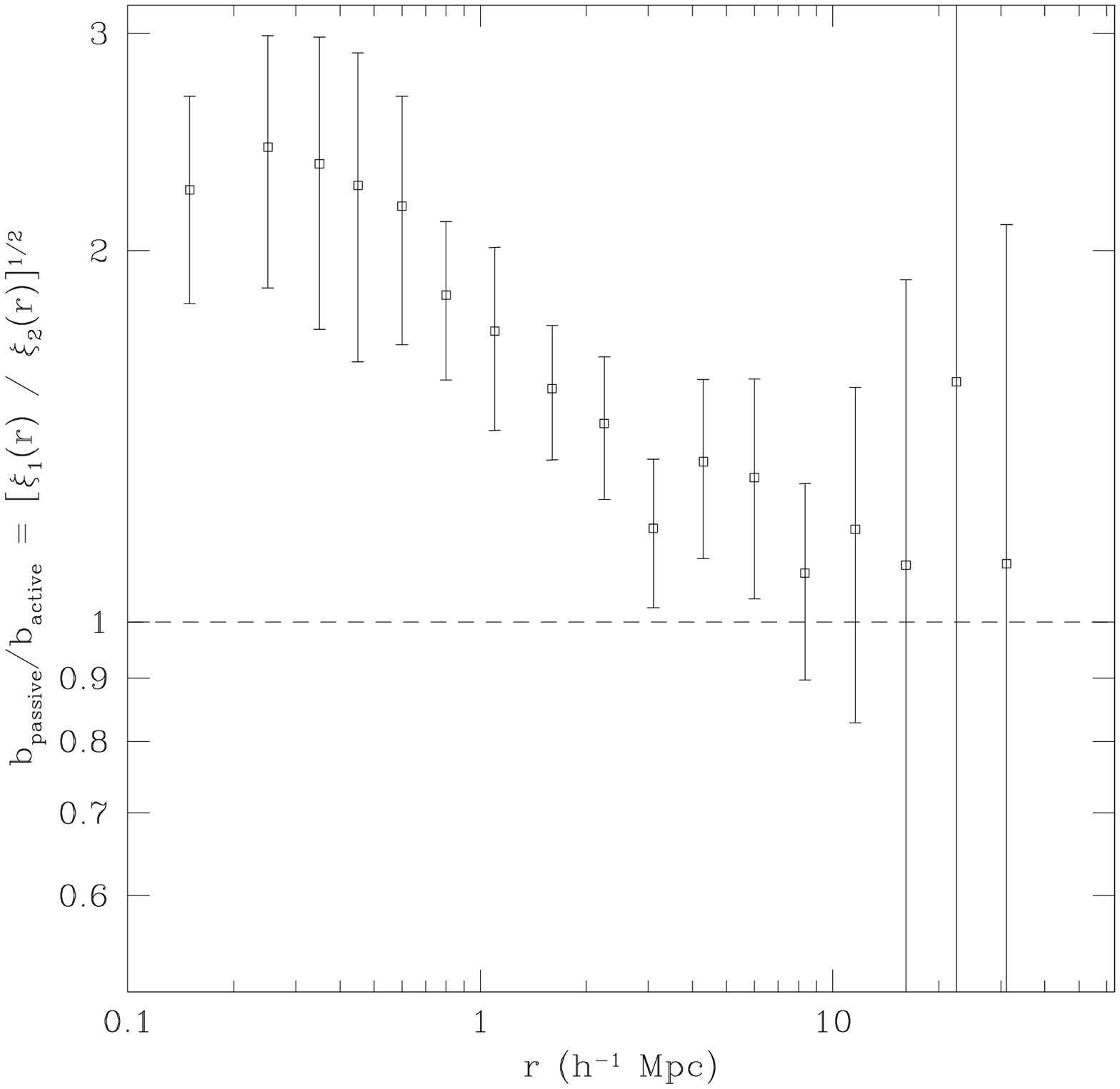}
\caption{Variations in the bias parameter with luminosity and spectral
type. The left panel shows the variation with luminosity of the galaxy
bias on a scale of $\sim$5\Mpc, relative to an $L^*$ galaxy (Norberg
\etal\ 2002a). The bias variations of the full 2dFGRS sample are
compared to subsamples with early and late spectral types, and to
earlier results by Norberg \etal\ (2001). The right panel shows the
relative bias of passive and actively star-forming galaxies as a
function of scale, over the range 0.2--20\Mpc\ (Madgwick \etal\ 2003).}
\label{2dFGRS_b}
\end{figure}

For the blue-selected 2dFGRS sample, it therefore seems that $L^*$
galaxies are nearly unbiased tracers of the low-redshift mass
distribution. However, this broad conclusion masks some very interesting
variations of the bias parameter with galaxy luminosity and type
(Fig.~\ref{2dFGRS_b}). Norberg \etal\ (2001, 2002a) show
conclusively that the bias parameter varies with luminosity, ranging
from $b = 1.5$ for bright galaxies to $b = 0.8$ for faint galaxies. The
relation between bias and luminosity is well represented by the simple
linear relation $b/b^* = 0.85 + 0.15 L/L^*$. They also find that, at all
luminosities, early-type galaxies have a higher bias than late-type
galaxies. A detailed comparison of the clustering of passive and
actively star-forming galaxies by Madgwick \etal\ (2003) shows that at
small separations, the passive galaxies cluster much more strongly, and
the relative bias ($b_{\rm passive}/b_{\rm active}$) is a decreasing
function of scale. On the largest scales, however, the relative bias
tends to a constant value of around 1.3.

\section{Redshift-space Distortions}
\label{2dFGRS_dist}

The redshift-space distortion of the clustering pattern can be modeled
as the combination of coherent infall on intermediate scales and random
motions on small scales. The compression of structures along the line of
sight due to coherent infall is quantified by the distortion parameter
$\beta \simeq \Omega^{0.6}/b$ (Kaiser 1987; Hamilton 1992). The random
motions are adequately modeled by an exponential distribution, $f(\upsilon) =
1/(a\sqrt{2})\exp(-\sqrt{2}|\upsilon|/a)$, where $a$ is the pairwise peculiar
velocity dispersion (also called $\sigma_{12}$).

The initial analysis of a subset of the 2dFGRS by Peacock \etal\ (2001)
obtained best-fit values of $\beta(L_s,z_s) = 0.43 \pm 0.07$ and $a =
385$ \kms\ at an effective weighted survey luminosity $L_s = 1.9L^*$ and
survey redshift $z_s = 0.17$. A more sophisticated reanalysis of the
full 2dFGRS by Hawkins \etal\ (2003) obtains $\beta(L_s,z_s) = 0.49 \pm
0.09$ and $a = 506 \pm 52$ \kms, with $L_s = 1.4L^*$ and $z_s = 0.15$
(right panel of Fig.~\ref{2dFGRS_stats}). These results, using different
fitting methods, are consistent, although the earlier result
underestimates the uncertainties by 20\%. Applying corrections based on
the variation in the bias parameter with luminosity and a constant
galaxy clustering model (Lahav \etal\ 2002) to the Hawkins \etal\ value
for the distortion parameter yields $\beta(L^*,z=0) = 0.47 \pm 0.08$.

\begin{figure}
%\centering
\hspace{-1.0cm}
\leavevmode
\includegraphics[clip,width=1.05\columnwidth]{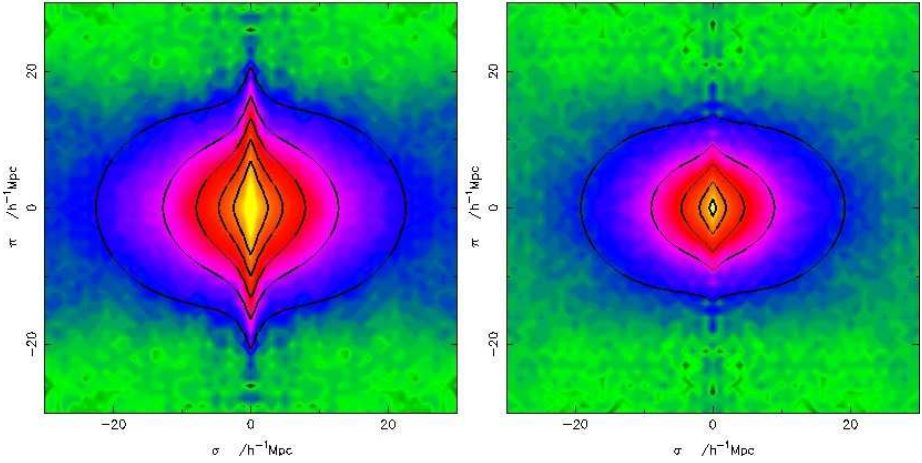}
\caption{The two-dimensional galaxy correlation function,
$\xi(\sigma,\pi)$, for passive (left) and actively star-forming (right)
galaxies (Madgwick \etal\ 2003). The grayscale image is the observed
$\xi(\sigma,\pi)$, and the contours show the best-fitting model.}
\label{2dFGRS_xi_type}
\end{figure}

Madgwick \etal\ (2003) extend this analysis to a comparison of the
active and passive galaxies, where the two-dimensional correlation
function, $\xi(\sigma,\pi)$, reveals differences in both the bias
parameter on large scales and the pairwise velocity dispersion on small
scales (Fig.~\ref{2dFGRS_xi_type}). The distortion parameter is
$\beta_{\rm passive} \simeq \Omega_m^{0.6}/b_{\rm passive} = 0.46 \pm
0.13$ for passive galaxies and $\beta_{\rm active} \simeq
\Omega_m^{0.6}/b_{\rm active} = 0.54 \pm 0.15$ for active galaxies; over
the range 8--20\Mpc\ the effective pairwise velocity dispersions are
$618 \pm 50$ \kms\ and $418 \pm 50$ \kms\ for passive and active galaxies, 
respectively.

\section{The Mass Density of the Universe}
\label{2dFGRS_omega}

The 2dFGRS provides a variety of ways to measure the mean mass density
of the Universe, along with the relative amounts of dark matter, baryons, 
and neutrinos.

\begin{figure}
%\centering
\hspace{-1.0cm}
\leavevmode
\includegraphics[clip,width=.50\columnwidth]{2dFGRS_mass_bar.eps} \hfil
\includegraphics[clip,width=.56\columnwidth]{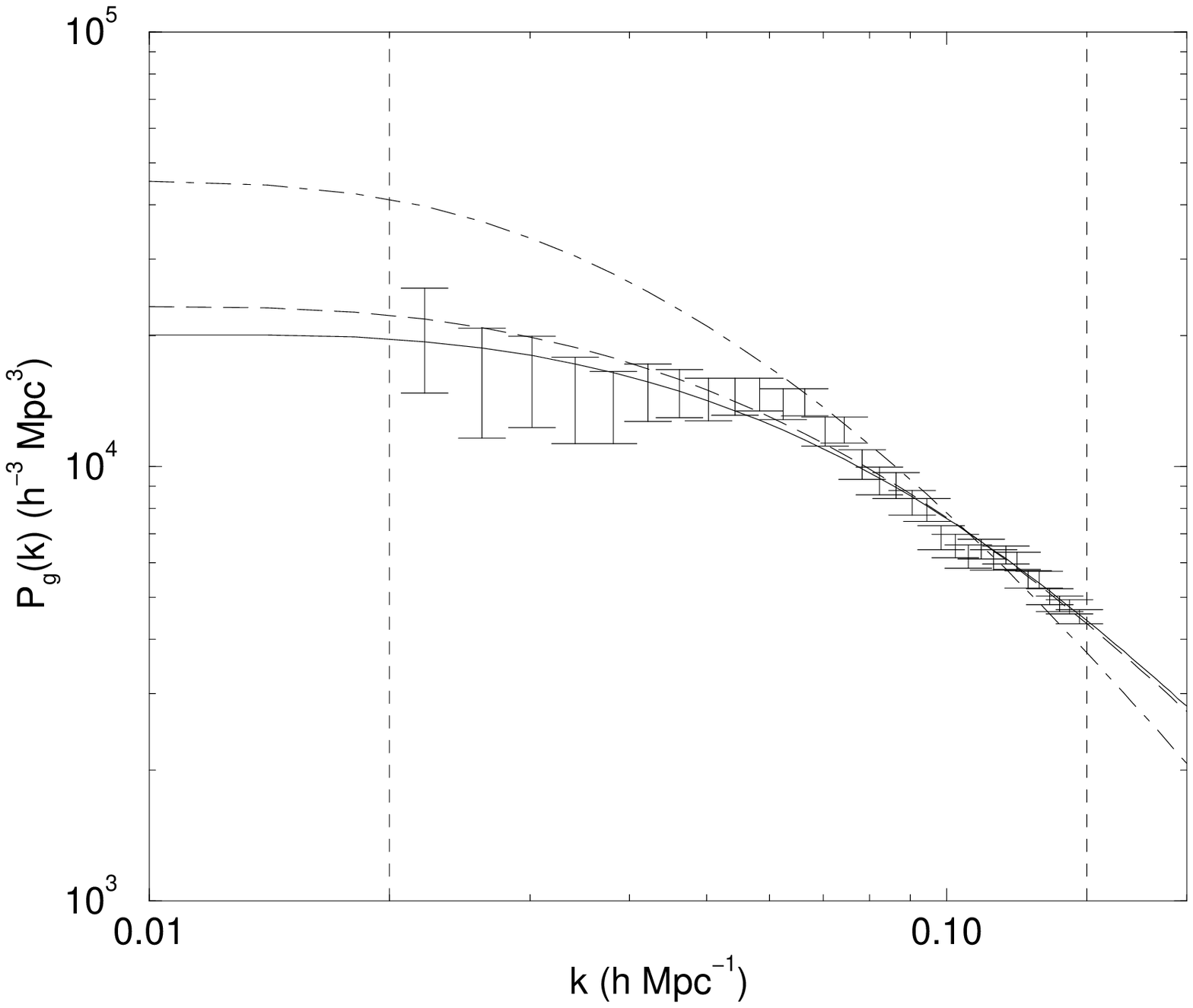}
\caption{Determinations of the mean mass density, $\Omega_m$, and the
baryon and neutrino mass fractions. The left panel shows the likelihood
surfaces obtained by fitting the full 2dFGRS power spectrum for the
shape parameter, $\Omega_m h$, and the baryon fraction,
$\Omega_b/\Omega_m$ (Peacock \etal\ 2003; cf.\ Percival \etal\ 2001).
The fit is over the well-determined linear regime ($0.02<k<0.15$\invMpc)
and assumes a prior on the Hubble constant of $h = 0.7 \pm 0.07$. The
right panel shows the fits to the 2dFGRS power spectrum (Elgar{\o}y \etal\
2002), assuming $\Omega_m=0.3$, $\Omega_{\Lambda}=0.7$, and $h=0.7$ for
three different neutrino densities: $\Omega_{\nu}$ = 0 (solid), 0.01
(dashed), and 0.05 (dot-dashed).}
\label{2dFGRS_mass}
\end{figure}

Fitting the shape of the galaxy power spectrum in the linear regime with
a model including both CDM and baryons (Percival \etal\
2001), and assuming that the Hubble constant is $h=0.7$ with a 10\%
uncertainty, yields a total mass density for the Universe of $\Omega_m =
0.29 \pm 0.07$ and a baryon fraction of $15\% \pm 7\%$ (i.e., $\Omega_b
= 0.044 \pm 0.021$). This analysis used 150,000 galaxies; a preliminary
reanalysis of the complete final sample of 221,000 galaxies with the
additional constraint that $n=1$ yields $\Omega_m = 0.26 \pm 0.05$ and
$\Omega_b = 0.044 \pm 0.016$ (Peacock \etal\ 2003; left panel of
Fig.~\ref{2dFGRS_mass}). Including neutrinos as a further constituent
of the mass allows an upper limit to be placed on their contribution to
the total density, based on the allowable degree of suppression of
small-scale structure due to the free streaming of neutrinos out of the
initial density perturbations (right panel of Fig.~\ref{2dFGRS_mass}).
Elgar{\o}y \etal\ (2002) obtain an upper limit on the neutrino mass
fraction of 13\% at the 95\% confidence level (i.e., $\Omega_{\nu} <
0.034$). This translates to an upper limit on the total neutrino mass
(summed over all species) of $m_{\nu} < 1.8 \, {\rm eV}$.

An alternative approach to deriving the total mass density is to use the
measurements in the quasi-linear regime of the redshift-space distortion
parameter $\beta \simeq \Omega_m^{0.6}/b$, in combination with estimates
of the bias parameter $b$ (Peacock \etal\ 2001; Hawkins \etal\ 2003).
Using the Lahav \etal\ (2002) estimate for $b$ gives $\Omega_m = 0.31
\pm 0.11$, while the Verde \etal\ (2002) value for $b$ gives $\Omega_m =
0.23 \pm 0.09$.

\section{Joint LSS-CMB Estimates of Cosmological Parameters}
\label{2dFGRS_joint}

Stronger constraints on these and other fundamental cosmological
parameters can be obtained by combining the power spectrum of the
present-day galaxy distribution from the 2dFGRS with the power spectrum
of the mass distribution at very early times derived from observations
of the anisotropies in the CMB. A general analysis of the combined CMB
and 2dFGRS data sets (Efstathiou \etal\ 2002) shows that, at the 95\%
confidence level, the Universe has a near-flat geometry ($\Omega_k
\approx 0 \pm 0.05$), with a low total matter density ($\Omega_m \approx
0.25 \pm 0.08$) and a large positive cosmological constant
($\Omega_{\Lambda} \approx 0.75 \pm 0.10$, consistent with the
independent estimates from observations of high-redshift supernovae). 

\begin{table}
\begin{center}
\leavevmode
\caption{Cosmological parameters from joint fits to the CMB and 2dFGRS
power spectra, assuming a flat geometry (Percival \etal\ 2002).}
\includegraphics[clip,width=.95\columnwidth]{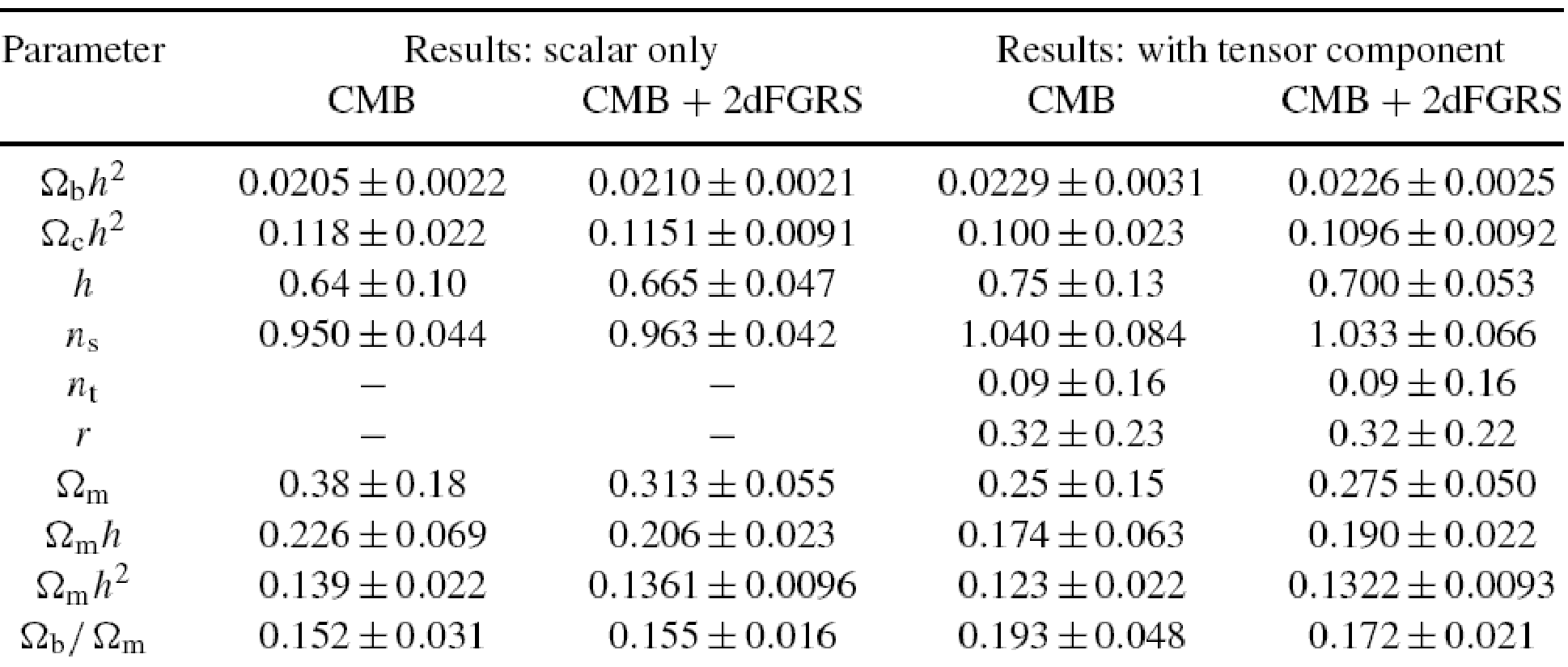}
\label{2dFGRS_lss_cmb}
\end{center} \vspace{-6pt}
{\footnotesize Note: the best-fit parameters and rms errors are obtained
by marginalizing over the likelihood distribution of the remaining
parameters. Results are given for scalar-only and scalar+tensor models,
and for the CMB power spectrum only and the CMB and 2dFGRS power spectra
jointly.}
\end{table}

If the models are limited to those with flat geometries (Percival \etal\
2002), then tighter constraints emerge (see Table~\ref{2dFGRS_lss_cmb}).
In this case the best estimate of the matter density is $\Omega_m = 0.31
\pm 0.06$, and the physical densities of CDM and baryons
are $\omega_c = \Omega_c h^2 = 0.12 \pm 0.01$ and $\omega_b = \Omega_b
h^2 = 0.022 \pm 0.002$; the latter agrees very well with the constraints
from Big Bang nucleosynthesis. This analysis also provides an estimate
of the Hubble constant ($H_0 = 67 \pm 5 \, {\rm km}\,  {\rm s}^{-1}\,  {\rm
Mpc}^{-1}$) that is independent of, but in excellent accord with, the
results from the {\it Hubble Space Telescope}\ Key Project. Comparing the
uncertainties on the various parameters in the CMB-only and CMB+2dFGRS
columns of Table~\ref{2dFGRS_lss_cmb} shows the very significant
improvements that are obtained by combining the CMB and 2dFGRS data sets.

Joint fits to the 2dFGRS and CMB power spectra also constrain the
equation of state parameter $w = p_{\rm vac}/\rho_{\rm vac}c^2$ for the
dark energy. Percival \etal\ (2002) find that in a flat Universe the
joint power spectra, together with the Hubble Key Project estimate for
$H_0$, imply an upper limit of $w < -0.52$ at the 95\% confidence level.

\section{The Galaxy Population}
\label{2dFGRS_gals}

Alongside these cosmological studies, the 2dFGRS has also produced a
wide range of results on the properties of the galaxy population, and
provided strong new constraints for models of galaxy formation and
evolution. Highlights in this area to date include: (1)~precise
determinations of the optical and near-IR galaxy luminosity functions
(Cole \etal\ 2001; Norberg \etal\ 2002b); (2)~a detailed
characterization of the variations in the luminosity function with
spectral type (Folkes \etal\ 1999; Madgwick \etal\ 2002); (3)~a
determination of the bivariate distribution of galaxies over luminosity
and surface brightness (Cross \etal\ 2001); (4)~a constraint on the
space density of rich clusters of galaxies from the velocity dispersion
distribution for identified clusters (De Propris \etal\ 2002);
(5)~separate radio luminosity functions for AGNs and star-forming
galaxies (Sadler \etal\ 2002; Magliocchetti \etal\ 2002);
(6)~constraints on the star formation history of galaxies from the mean
spectrum of galaxies in the local Universe (Baldry \etal\ 2002); (7)~a
measurement of the environmental dependence of star formation rates of
galaxies in clusters (Lewis \etal\ 2002b); and (8)~a comparison of the
field and cluster luminosity functions for galaxies with difference
spectral types (De Propris \etal\ 2003).

\begin{figure}
%\centering
\hspace{-1.0cm}
\leavevmode
\includegraphics[clip,width=1.05\columnwidth]{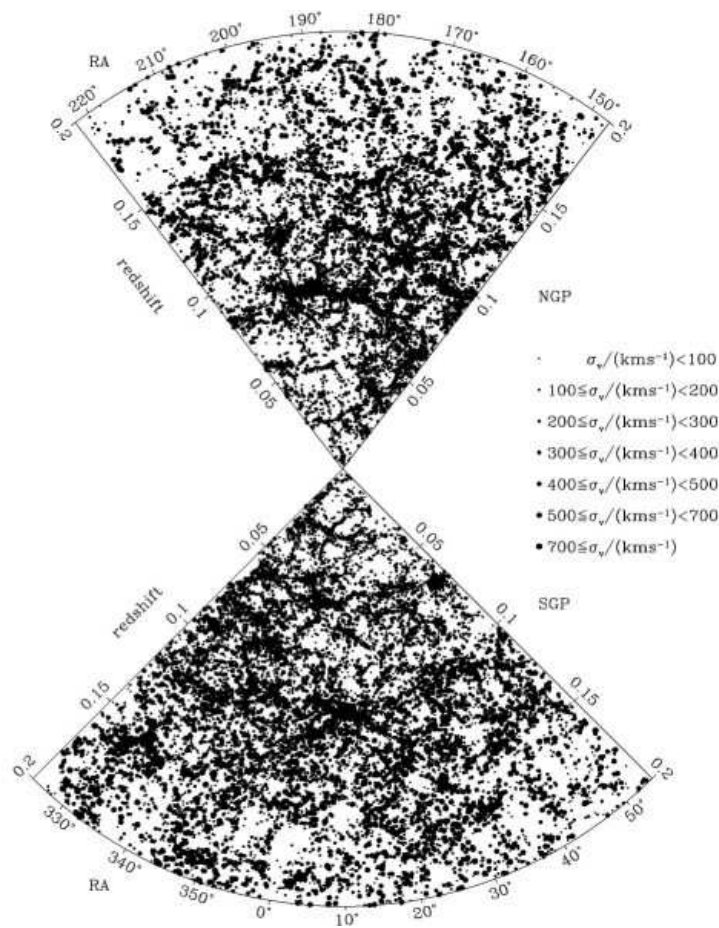}
\caption{A redshift slice showing the distribution of groups and
clusters identified within the 2dFGRS using a three-dimensional
friends-of-friends algorithm in position and redshift space (Eke \etal\
2003). The number of members found in each cluster is shown by different
gray shading; the estimated velocity dispersion is indicated by the size
of the dot.}
\label{2dFGRS_groups_slice}
\end{figure}

The next step will be to further investigate the correlations between
these properties and the local environment of each galaxy, quantified
through the local galaxy density or the new group and cluster catalog
that has been constructed from the positions and velocity information in
the 2dFGRS (Eke \etal\ 2003; see Fig.~\ref{2dFGRS_groups_slice}).

\section{Conclusions}
\label{2dFGRS_conc}

The measurement of cosmological parameters from the 2dFGRS has made a
significant contribution to shaping the current consensus model for the
fundamental properties of the Universe that has emerged from a range of
independent observations, including the measurements of the CMB
anisotropies, the distances to high-redshift supernovae, and Big Bang
nucleosynthesis. The results obtained to date only represent a fraction
of the information that can be extracted from the 2dFGRS on the
properties of galaxies and their relation to the large-scale structure
of the galaxy distribution. Much more is still to emerge from analysis
of the survey, and from combining the 2dFGRS with other large surveys
and with detailed follow-up observations.

Further information on the 2dF Galaxy Redshift Survey can be found on
the WWW at http://www.mso.anu.edu.au/2dFGRS.

\vspace{0.3cm}
{\bf Acknowledgements}.
These results are presented on behalf of the 2dFGRS team: Ivan K.\
Baldry, Carlton M.\ Baugh, Joss Bland-Hawthorn, Sarah Bridle, Terry
Bridges, Russell Cannon, Shaun Cole, Matthew Colless, Chris Collins,
Warrick Couch, Nicholas Cross, Gavin Dalton, Roberto De Propris, Simon
P.\ Driver, George Efstathiou, Richard S.\ Ellis, Carlos S.\ Frenk, Karl
Glazebrook, Edward Hawkins, Carole Jackson, Bryn Jones, Ofer Lahav, Ian
Lewis, Stuart Lumsden, Steve Maddox, Darren Madgwick, Peder Norberg,
John A.\ Peacock, Will Percival, Bruce A.\ Peterson, Will Sutherland, and
Keith Taylor. The 2dFGRS was made possible through the dedicated efforts
of the staff of the Anglo-Australian Observatory, both in creating the
2dF instrument and in supporting it on the telescope.


\begin{thereferences}{}

\bibitem{}
Baldry, I., \etal\ 2002, \apj, 569, 582

\bibitem{}
Cole, S., \etal\ 2001, \mnras, 326, 255

\bibitem{}
Colless, M. M. 1999, Phil. Trans. R. Soc. Lond. A, 357, 105

\bibitem{}
Colless, M. M., \etal\ 2001, \mnras, 328, 1039

\bibitem{}
Cross, N., \etal\ 2001, \mnras, 324, 825

\bibitem{}
De Propris, R., \etal\ 2002, \mnras, 329, 87

\bibitem{}
------. 2003, \mnras, submitted (astro-ph/0212562)

\bibitem{}
Efstathiou, G., \etal\ 2002, \mnras, 330, L29

\bibitem{}
Eke, V., \etal\ 2003, in preparation

\bibitem{}
Elgar{\o}y, \O., \etal\ 2002, Phys. Rev. Lett., 89, 061301

\bibitem{}
Folkes, S., \etal\ 1999, \mnras, 308, 459

\bibitem{}
Hamilton, A.~J.~S. 1992, \apj, 385, L5

\bibitem{}
Hawkins, E., \etal\ 2003, \mnras, submitted (astro-ph/0212375)

\bibitem{}
Kaiser, N. 1987, \mnras, 227, 1

\bibitem{}
Lahav, O., \etal\ 2002, \mnras, 333, 961

\bibitem{}
Lewis, I., \etal\ 2002a, \mnras, 333, 279

\bibitem{}
------. 2002b, \mnras, 334, 673

\bibitem{}
Maddox, S.~J., Efstathiou, G., Sutherland, W. J., \& Loveday, J. 1990, \mnras, 
242, 43P

\bibitem{}
Madgwick, D. S., \etal\ 2002, MNRAS, 333, 133

\bibitem{}
------. 2003, in preparation

\bibitem{}
Magliocchetti, M., \etal\ 2002, \mnras, 333, 100

\bibitem{}
Norberg, P., \etal\ 2001, \mnras, 328, 64

\bibitem{}
------. 2002a, \mnras, 332, 827

\bibitem{}
------. 2002b, \mnras, 336, 907

\bibitem{}
Peacock, J. A., \etal\ 2001, Nature, 410, 169

\bibitem{}
------. 2003, in preparation

\bibitem{}
Percival, W. J., \etal\ 2001, \mnras, 327, 1297

\bibitem{}
------. 2002, \mnras, 337, 1068

\bibitem{}
Sadler, E.~M., \etal\ 2002, \mnras, 329, 227

\bibitem{}
Saunders, W., \etal\ 2000, \mnras, 317, 55

\bibitem{}

Shectman, S.~A., Landy, S.~D., Oemler, A., Tucker, D.~L., Lin, H., Kirshner,
R.~P., \& Schechter, P.~L. 1996, \apj, 470, 172

\bibitem{}
Taylor, K., \& Gray P. 1990, Proc.\ SPIE, 1236, 290

\bibitem{}
Verde, L., \etal\ 2002, \mnras, 335, 432
\end{thereferences}
\end{document}